\documentclass[10pt]{article}
\usepackage{multicol}
\usepackage{graphicx}
\usepackage{amsmath}
\usepackage[a4paper]{geometry}

\begin{document}
\begin{center}
{\Large \bf Comparing a few distributions of transverse momenta\\
in high energy collisions}

\vskip.75cm

Qi Wang, Pei-Pin Yang, Fu-Hu Liu{\footnote{E-mail:
fuhuliu@163.com; fuhuliu@sxu.edu.cn}}

\vskip.25cm

{\small\it Institute of Theoretical Physics \& State Key
Laboratory of Quantum Optics and Quantum Optics Devices,\\ Shanxi
University, Taiyuan, Shanxi 030006, China}
\end{center}

\vskip.5cm

{\bf Abstract:} Transverse momentum spectra of particles produced
in high energy collisions are very important due to their
relations to the excitation degree of interacting system. To
describe the transverse momentum spectra, one can use more than
one probability density functions of transverse momenta, which are
simply called the functions or distributions of transverse momenta
in some cases. In this paper, a few distributions of transverse
momenta in high energy collisions are compared with each other in
terms of plots to show some quantitative differences. Meanwhile,
in the framework of Tsallis statistics, the distributions of
momentum components, transverse momenta, rapidities, and
pasudorapidities are obtained according to the analytical and
Monte Carlo methods. These analyses are useful to understand
carefully different distributions in high energy collisions.
\\

\\


\vskip1.0cm

\begin{multicols}{2}

{\section{Introduction}}

In high energy hadron-hadron, hadron-nucleus, and nucleus-nucleus
collisions, transverse momentum spectra of secondary particles are
one of the ``first day" measurable quantities. The transverse
momentum spectra are expected to reflect the excitation degree of
interacting system, which is useful to understand the properties
of particle production and system evolution. To describe the
transverse momentum spectra, one can use more than one probability
density functions of transverse momenta. Strictly speaking, the
probability density function and the distribution function are
different concepts in statistics. We simply call the probability
density function the distribution or function in some cases in the
present work.

Three types of distributions will be compared with respective
modified forms in the present work. Firstly, we compare the
Hagedorn function with its modified forms which are suitable to
fit the spectra in high transverse momentum region. Secondly, we
compare the simplest standard distribution with its modified forms
which are suitable to fit the spectra in low transverse momentum
region. Thirdly, we compare the simplest Boltzmann distribution
with its modified forms which are also suitable to fit the spectra
in low transverse momentum region, though the Boltzmann
distribution is one of the standard distribution.

After comparisons for transverse momentum distributions, we
discuss an application of the Monte Carlo method according to the
(transverse) momentum distribution and the assumption of isotropic
emission in the subsequent part of the present work. In
particular, in the framework of Tsallis statistics, the
distributions of momentum components, transverse momenta,
rapidities, and pasudorapidities are obtained according to the
analytical and Monte Carlo methods.
\\

{\section{Formalism and method}}

i) {\it The Hagedorn function and its modified forms}

The Hagedorn function and its modified forms are suitable to
describe the transverse momentum ($p_T$) spectra of heavy flavor
particles which are expectantly produced from the hard scattering
process and distributed usually in a wider $p_T$ range. In
general, the wider $p_T$ range is from 0 to the maximum $p_T$.

In refs. [1, 2], an inverse power-law
\begin{align}
f_1(p_T)=\frac{1}{N}\frac{dN}{dp_T}=A_1p_T\bigg(1+\frac{p_T}{p_1}
\bigg)^{-n_1}
\end{align}
that is an empirical formula inspired by quantum chromodynamics
(QCD) is used, where $N$ denotes the number of particles, $p_1$
and $n_1$ are the free parameters, and $A_1$ is the normalization
constant. We call this type of inverse power-law the Hagedorn
function [1].

In ref. [3], a modified Hagedorn function is shown as
\begin{align}
f_2(p_T)=\frac{1}{N}\frac{dN}{dp_T}=A_2\frac{p^2_T}{\sqrt{p_T^2+m_0^2}}
\bigg(1+\frac{p_T}{p_2} \bigg)^{-n_2},
\end{align}
where $m_0$ is the rest mass of considered particle, $p_2$ and
$n_2$ are the free parameters, and $A_2$ is the normalization
constant. We call Eq. (2) the first modified Hagedorn function.

In ref. [4--8], there is another inverse power-law
\begin{align}
f_3(p_T)=\frac{1}{N}\frac{dN}{dp_T}=A_3p_T\bigg[1+\bigg(\frac{p_T}{p_3}\bigg)^2
\bigg]^{-n_3},
\end{align}
where $p_3$ and $n_3$ are the free parameters, and $A_3$ is the
normalization constant. We call Eq. (3) the second modified
Hagedorn function.

Even in ref. [9], there is the form
\begin{align}
f_4(p_T)=\frac{1}{N}\frac{dN}{dp_T}=A_4\bigg[1+\bigg(\frac{p_T}{p_4}\bigg)^2
\bigg]^{-n_4},
\end{align}
where $p_4$ and $n_4$ are the free parameters, and $A_4$ is the
normalization constant. We call Eq. (4) the third modified
Hagedorn function.
\\

ii) {\it The simplest standard distribution and its modified
forms}

The simplest standard distribution and its modified forms are
suitable to describe the $p_T$ spectra of light flavor particles
which are expectantly produced from the soft excitation process or
thermal process and distributed mainly in a narrow $p_T$ range.
The narrow $p_T$ range covers a range from 0 to around 2$\sim$3
GeV/$c$ for pions produced in collisions at dozes of GeV. The
boundary of narrow $p_T$ range is changeable for different
particles and at different energies.

The standard distribution has different forms. In the case of
including rapidity and chemical potential, the simplest form can
be written as [10]
\begin{align}
f_1(p_T)=& \frac{1}{N}\frac{dN}{dp_T}= C_1p_T \times \nonumber\\
& \int^{y_{\max}}_{y_{\min}} \bigg[\exp
\bigg(\frac{\sqrt{p_T^2+m_0^2} \cosh y-\mu}{T_1} \bigg)+S
\bigg]^{-1}dy,
\end{align}
where $y_{\min}$ and $y_{\max}$ denote the minimum and maximum $y$
respectively, $\mu$ denotes the chemical potential, $T_1$ is the
free parameter of temperature, and $C_1$ is the normalization
constant. In particular, $S=1$, 0, and $-1$ denote the
Fermi-Dirac, Maxwell-Boltzmann, and Bose-Einstein statistics,
respectively. This form is inconsistent with the classical ideal
gas model, though it has many applications.

A modified form of the simplest standard distribution is [10, 11]
\begin{align}
f_2(p_T)=& \frac{1}{N}\frac{dN}{dp_T}= C_2p_T\sqrt{p_T^2+m_0^2}
\int^{y_{\max}}_{y_{\min}} \cosh y \times \nonumber\\
& \bigg[\exp \bigg(\frac{\sqrt{p_T^2+m_0^2} \cosh y-\mu}{T_2}
\bigg)+S \bigg]^{-1}dy,
\end{align}
where $T_2$ is the free parameter of temperature and $C_2$ is the
normalization constant. We call Eq. (6) the first modified the
simplest standard distribution. This form is consistent with the
classical ideal gas model, i.e. it is close to Rayleigh
distribution at low energy.

Another modified form of the simplest standard distribution is
[12]
\begin{align}
f_3(p_T)=& \frac{1}{N}\frac{dN}{dp_T}= C_3p_T^2 \times \nonumber\\
& \int^{y_{\max}}_{y_{\min}} \bigg[\exp
\bigg(\frac{\sqrt{p_T^2+m_0^2} \cosh y-\mu}{T_3} \bigg)+S
\bigg]^{-1}dy,
\end{align}
where $T_3$ is the free parameter of temperature and $C_3$ is the
normalization constant. We call Eq. (7) the second modified the
simplest standard distribution. This form is also inconsistent
with the classical ideal gas model.
\\

iii) {\it The simplest Boltzmann distribution and its modified
forms}

In some cases, we can neglect chemical potential and/or spin
effect, and/or consider only mid-rapidity, in the simplest
standard distribution and its modified forms. In the case of
neglecting simultaneously chemical potential and spin effect, and
considering only mid-rapidity, we have simpler forms of the above
Eqs. (5)--(7) to be
\begin{align}
f_1(p_T)=& \frac{1}{N}\frac{dN}{dp_T}= C_1p_T \exp \bigg(-
\frac{\sqrt{p_T^2+m_0^2}}{T_1} \bigg),
\end{align}
\begin{align}
f_2(p_T)& = \frac{1}{N}\frac{dN}{dp_T} \nonumber\\
&= C_2p_T\sqrt{p_T^2+m_0^2} \exp \bigg(-
\frac{\sqrt{p_T^2+m_0^2}}{T_2} \bigg),
\end{align}
and
\begin{align}
f_3(p_T)=& \frac{1}{N}\frac{dN}{dp_T}= C_3p_T^2 \exp \bigg(-
\frac{\sqrt{p_T^2+m_0^2}}{T_3} \bigg),
\end{align}
respectively. We call Eqs. (9) and (10) the first and second
modified the simplest Boltzmann distribution respectively. Only
Eq. (9) is consistent with the classical ideal gas model at low
energy.

It should be noted that, although the same symbols are used in
different functions, they have different values in general. In
some cases, the differences are larger due to different
interactions and processes.
\\

iv) {\it Monte Carlo calculation based on $p_T$ distribution}

Based on one of $p_T$ distributions and the assumption of
isotropic emission, we can obtain other distributions. In
particular, if the analytic expression is difficult to obtain, we
can use the Monte Carlo method to obtain some concerned quantities
and their distributions.

In the Monte Carlo method [13], let $R_{1,2,3}$ denote random
numbers distributed evenly in [0,1]. Some discrete values of $p_T$
can be obtained due to the following limitation
\begin{align}
\int_0^{p_T}f_{p_T}(p'_T)dp'_T <R_1 <\int_0^{p_T+\delta
p_T}f_{p_T}(p'_T) dp'_T,
\end{align}
where $\delta p_T$ denote a small shift relative to $p_T$.

Under the assumption of isotropic emission in the rest frame, we
have the momentum components to be
\begin{align}
p_x=p_T\cos\phi, \hskip2mm p_y=p_T\sin\phi, \hskip2mm
p_z=p_T/\tan\theta,
\end{align}
where
\begin{align}
\phi= 2\pi R_2, \hskip2mm \theta=2\arcsin\sqrt{R_3}
\end{align}
due to $\phi$ and $\theta$ satisfy the distributions
\begin{align}
f_{\phi}(\phi)=\frac{1}{2\pi}, \hskip2mm
f_{\theta}(\theta)=\frac{1}{2}\sin\theta
\end{align}
respectively [13].

The momentum $p$ and energy $E$ can be obtained by
\begin{align}
p=\sqrt{p_z^2+p_T^2}, \hskip2mm E=\sqrt{p^2+m_0^2}.
\end{align}
Further, the velocity components are
\begin{align}
\beta_{x}=\frac{p_{x}}{E}, \hskip2mm \beta_{y}=\frac{p_{y}}{E},
\hskip2mm \beta_{z}=\frac{p_{z}}{E}.
\end{align}
The rapidity $y$ and pseudorapidity $\eta$ are [14]
\begin{align}
y\equiv\frac{1}{2} \ln \bigg(\frac{E+p_z}{E-p_z} \bigg), \hskip2mm
\eta\equiv-\ln \tan\bigg( \frac{\theta}{2}\bigg).
\end{align}
According to definition of $y$, we can define $y_1$ by $E$ and
$p_x$, and $y_2$ by $E$ and $p_y$, to be
\begin{align}
y_1\equiv\frac{1}{2} \ln \bigg(\frac{E+p_x}{E-p_x} \bigg),
\hskip2mm y_2\equiv\frac{1}{2} \ln \bigg(\frac{E+p_y}{E-p_y}
\bigg).
\end{align}

Combining with the distribution of $\theta$ and the definition of
$\eta$, we have the distribution of $\eta$ to be
\begin{align}
f_{\eta}(\eta)=\frac{1}{2\cosh^2\eta}
\end{align}
which satisfies approximately the Gaussian distribution with the
width of $\sigma_{\eta}\approx0.91$ [15]. The distribution of $y$
is expected to obey the Gaussian distribution with the width of
$\sigma_y<\sigma_{\eta}$.

According to $p_T$ distribution and isotropic assumption, many
quantities can be obtained. In fact, the scatter plots of
particles in the three-dimensional momentum ($p_x$-$p_y$-$p_z$),
velocity ($\beta_x$-$\beta_y$-$\beta_z$), and rapidity
($y_1$-$y_2$-$y$) spaces can be obtained based on the above
discussions. We shall not discuss the scatter plots of particles
due to they being beyond the focus of the present work.
\\

v) {\it Analytical and Monte Carlo calculations based on momentum
distribution}

Although we can obtain other distributions based on $p_T$
distributions and the assumption of isotropic emission, consistent
$p_T$ and $y$ distributions should be obtained from the momentum
($p$) distribution and the assumption of isotropic emission. There
are various $p$ distributions which may be from the Fermi-Dirac,
Maxwell-Boltzmann, Bose-Einstein, or Tsallis statistics, etc. As
an example, we use the the $p$ distribution in the Tsallis
statistics.

In the Tsallis statistics, one has [10, 16--18]
\begin{align}
f_p(p)&=\frac{1}{N}\frac{dN}{dp}
=Cp^2\bigg[1+\frac{q-1}{T}\sqrt{p^2+m_0^2}\bigg]^{-\frac{q}{q-1}},
\end{align}
where $T$ is the temperature, $q$ is the entropy index, $C$ is the
normalization constant, and $\mu$ and $S$ are neglected for
convenient treatment. The invariant $p$ distribution is
\begin{align}
E\frac{d^3N}{dp^3}=C\sqrt{p^2+m_0^2}
\bigg[1+\frac{q-1}{T}\sqrt{p^2+m_0^2}\bigg]^{-\frac{q}{q-1}}.
\end{align}

The distributions of unit $p_T$ and $y$, $p_T$, $y$, and $p_x$ are
\begin{align}
f_{p_T,y}(p_T,y)=&\frac{1}{N}\frac{d^2N}{dp_Tdy}=
Cp_T\sqrt{p_T^2+m_0^2} \cosh y \times
\nonumber\\
& \bigg[1+\frac{q-1}{T}\sqrt{p_T^2+m_0^2} \cosh y
\bigg]^{-\frac{q}{q-1}},
\end{align}
\begin{align}
f_{p_T}(p_T)=& \frac{1}{N} \frac{dN}{dp_T}= Cp_T\sqrt{p_T^2+m_0^2} \int_{y_{\min}}^{y_{\max}}\cosh y \times \nonumber\\
& \bigg[1+\frac{q-1}{T}\sqrt{p_T^2+m_0^2} \cosh y
\bigg]^{-\frac{q}{q-1}}dy,
\end{align}
\begin{align}
f_y(y)=& \frac{1}{N} \frac{dN}{dy}= C\cosh y \int_0^{\infty}
p_T\sqrt{p_T^2+m_0^2} \times
\nonumber\\
& \bigg[1+\frac{q-1}{T}\sqrt{p_T^2+m_0^2} \cosh y
\bigg]^{-\frac{q}{q-1}}dp_T,
\end{align}
\begin{align}
f_{p_x}(p_x)=& \frac{1}{N} \frac{dN}{dp_x} \nonumber\\
=& \frac{1}{2\pi} \int_{-\infty}^{\infty}
\frac{1}{\sqrt{p_x^2+p_y^2}} f_{p_T}\left(\sqrt{p_x^2+p_y^2}
\right)dp_y \nonumber\\
=& \frac{C}{2\pi}\int_{-\infty}^{\infty} \sqrt{p_x^2+p_y^2+m_0^2} \int_{y_{\min}}^{y_{\max}}\cosh y \times \nonumber\\
& \bigg[1+\frac{q-1}{T}\sqrt{p_x^2+p_y^2+m_0^2} \cosh y
\bigg]^{-\frac{q}{q-1}}dydp_y,
\end{align}
respectively, where $C$ in the above equations may be different
from each other.

In the Monte Carlo method [13], some discrete values of $p$ can be
obtained due to the following limitation
\begin{align}
\int_0^{p}f_{p}(p')dp' <R_1 <\int_0^{p+\delta p}f_{p}(p') dp',
\end{align}
where $\delta p$ denote a small shift relative to $p$. Then
\begin{align}
p_T=p\sin\theta=p\sin\left(2\arcsin \sqrt{R_3} \right).
\end{align}
Other quantities have the same expressions as those in subsection
iv).
\\

{\section{Results and discussion}}

\begin{figure*}
\begin{center}
\includegraphics[width=7.0cm]{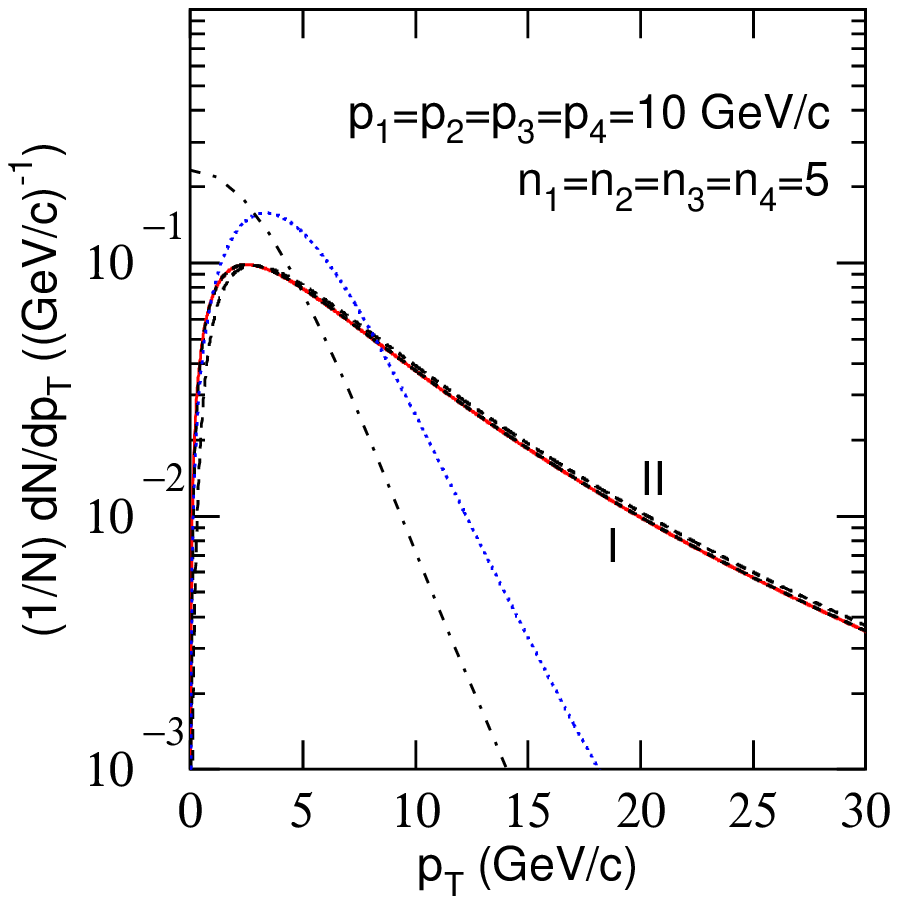}
\end{center}
{\small Fig. 1. Transverse momentum spectra obtained from the
Hagedorn function and its modified forms. The solid, dashed,
dotted, and dot-dashed curves represent the results from Eqs.
(1)--(4), respectively, with $p_1=p_2=p_3=p_4=10$ GeV/$c$ and
$n_1=n_2=n_3=n_4=5$. In particular, the dashed curves with marks I
and II corresponding to $m_0=0.139$ and 0.938 GeV/$c^2$ in Eq. (2)
respectively.}
\end{figure*}

\begin{figure*}
\begin{center}
\includegraphics[width=12cm]{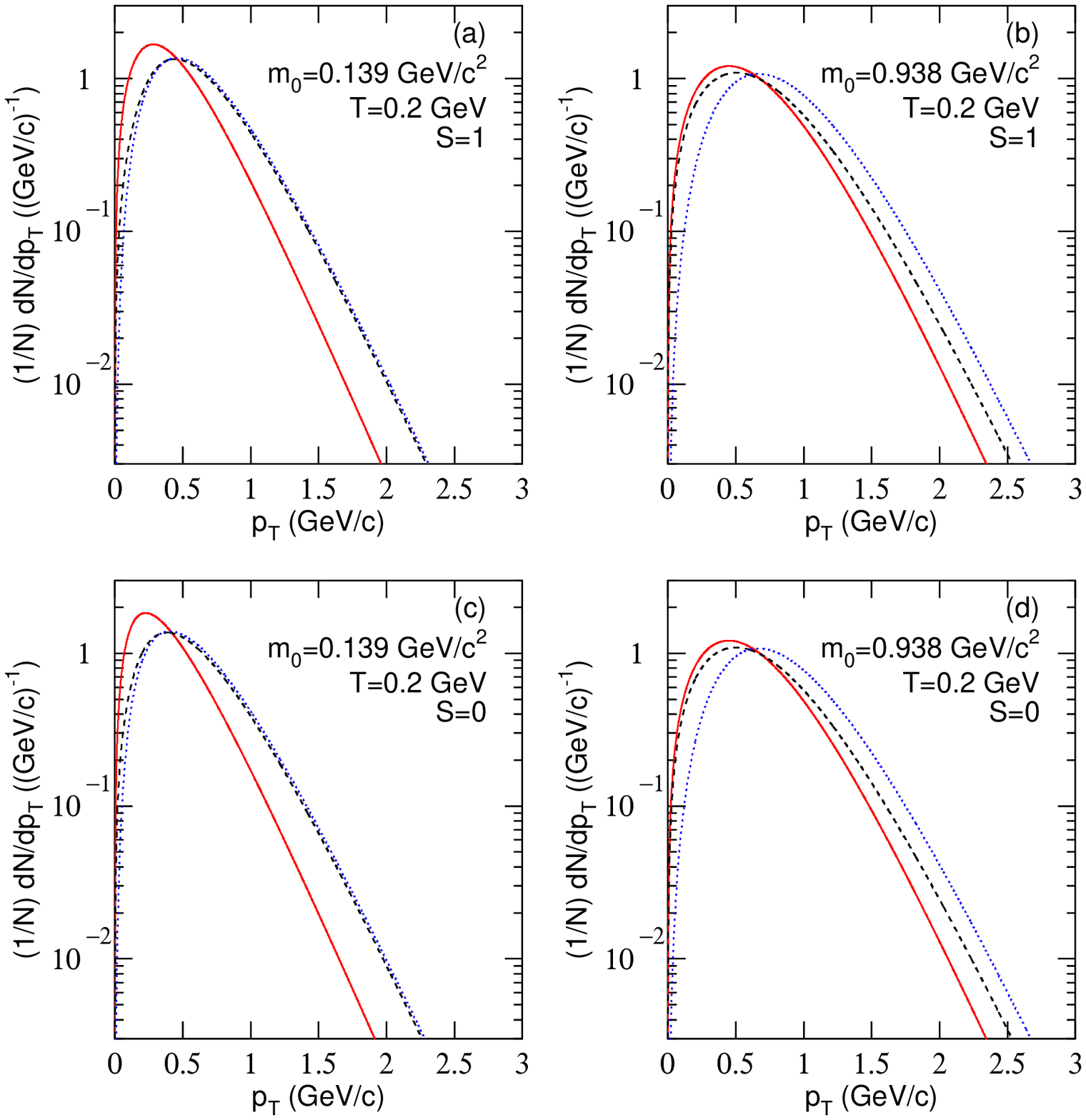}
\vskip.35cm
\includegraphics[width=12cm]{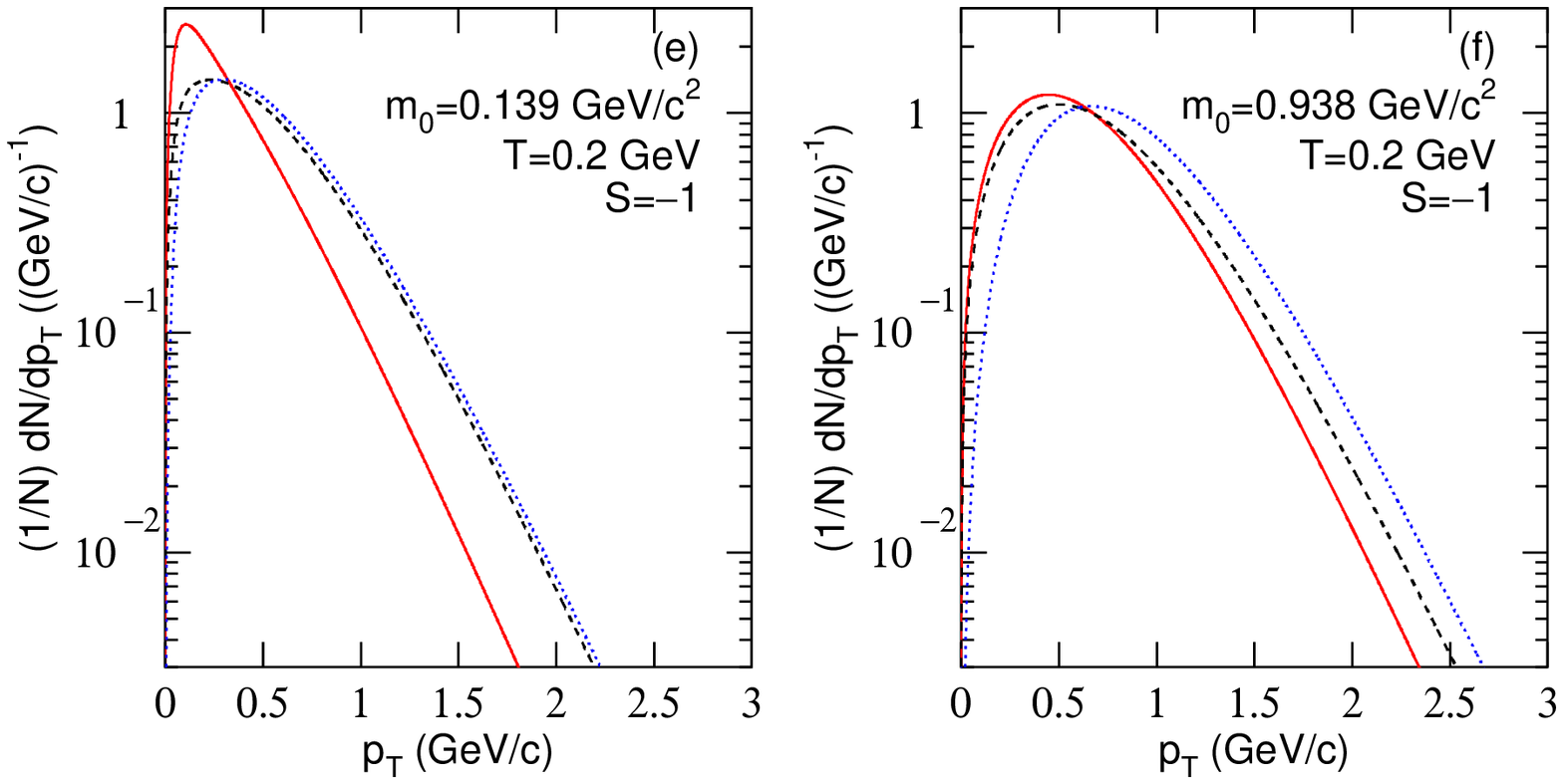}
\end{center}
{\small Fig. 2. Transverse momentum spectra obtained from the
simplest standard distribution and its modified forms. The solid,
dashed, and dotted curves represent the results from Eqs.
(5)--(7), respectively, with $T_1=T_2=T_3=T=0.2$ GeV, $\mu=0.1$
GeV, $y_{\min}=-0.5$, and $y_{\max}=0.5$. Panels (a)(b), (c)(d),
and (e)(f) correspond to $S=1$, 0, and $-1$, respectively; and
panels (a)(c)(e) and (b)(d)(f) correspond to $m_0=0.139$ and 0.938
GeV/$c^2$ respectively.}
\end{figure*}

\begin{figure*}
\begin{center}
\includegraphics[width=12cm]{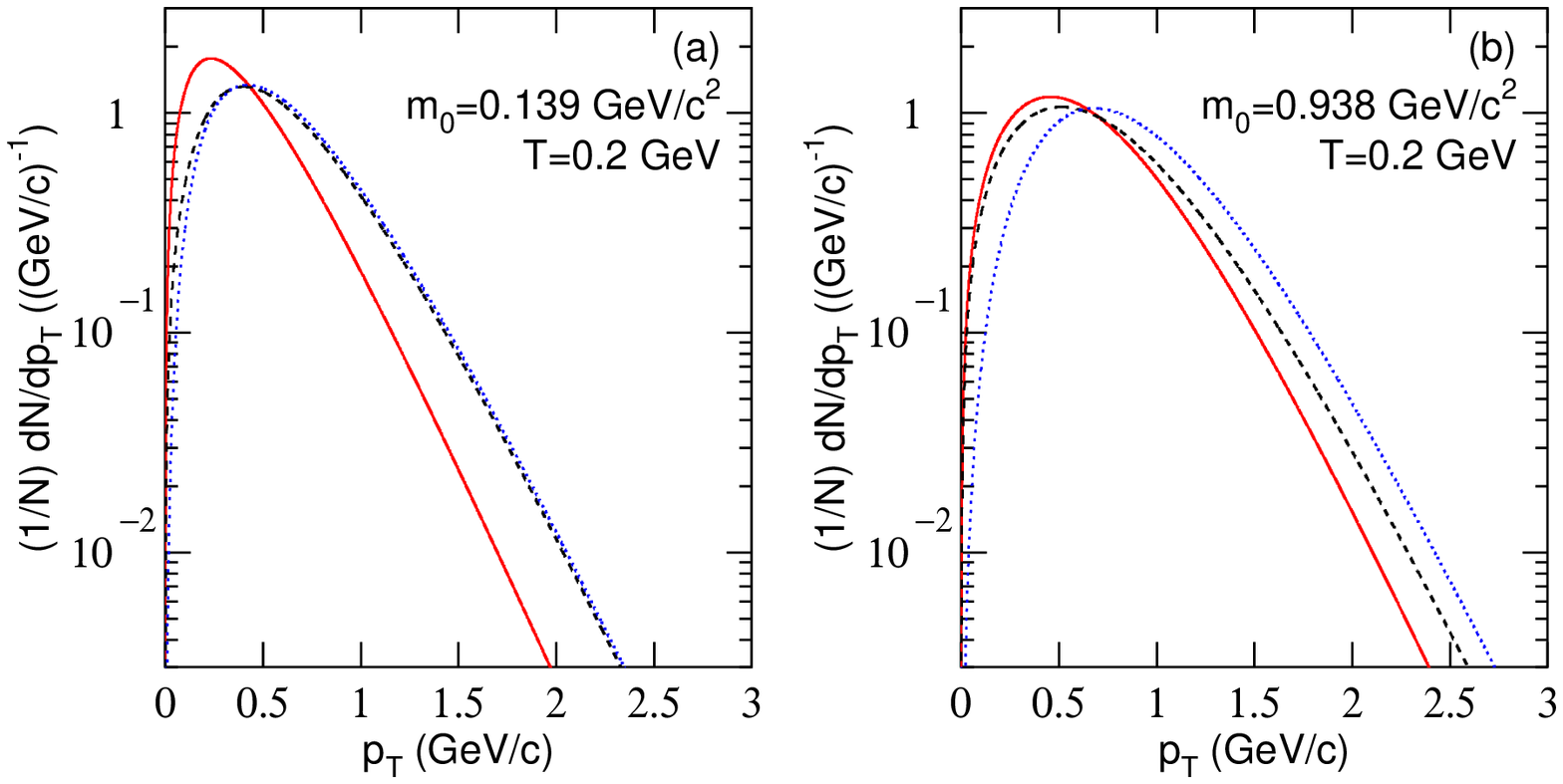}
\end{center}
{\small Fig. 3. Transverse momentum spectra obtained from the
simplest Boltzmann distribution and its modified forms. The solid,
dashed, and dotted curves represent the results from Eqs.
(8)--(10), respectively, with $T_1=T_2=T_3=T=0.2$ GeV. Panels (a)
and (b) correspond to $m_0=0.139$ and 0.938 GeV/$c^2$
respectively.}
\end{figure*}

\begin{figure*}
\begin{center}
\includegraphics[width=12cm]{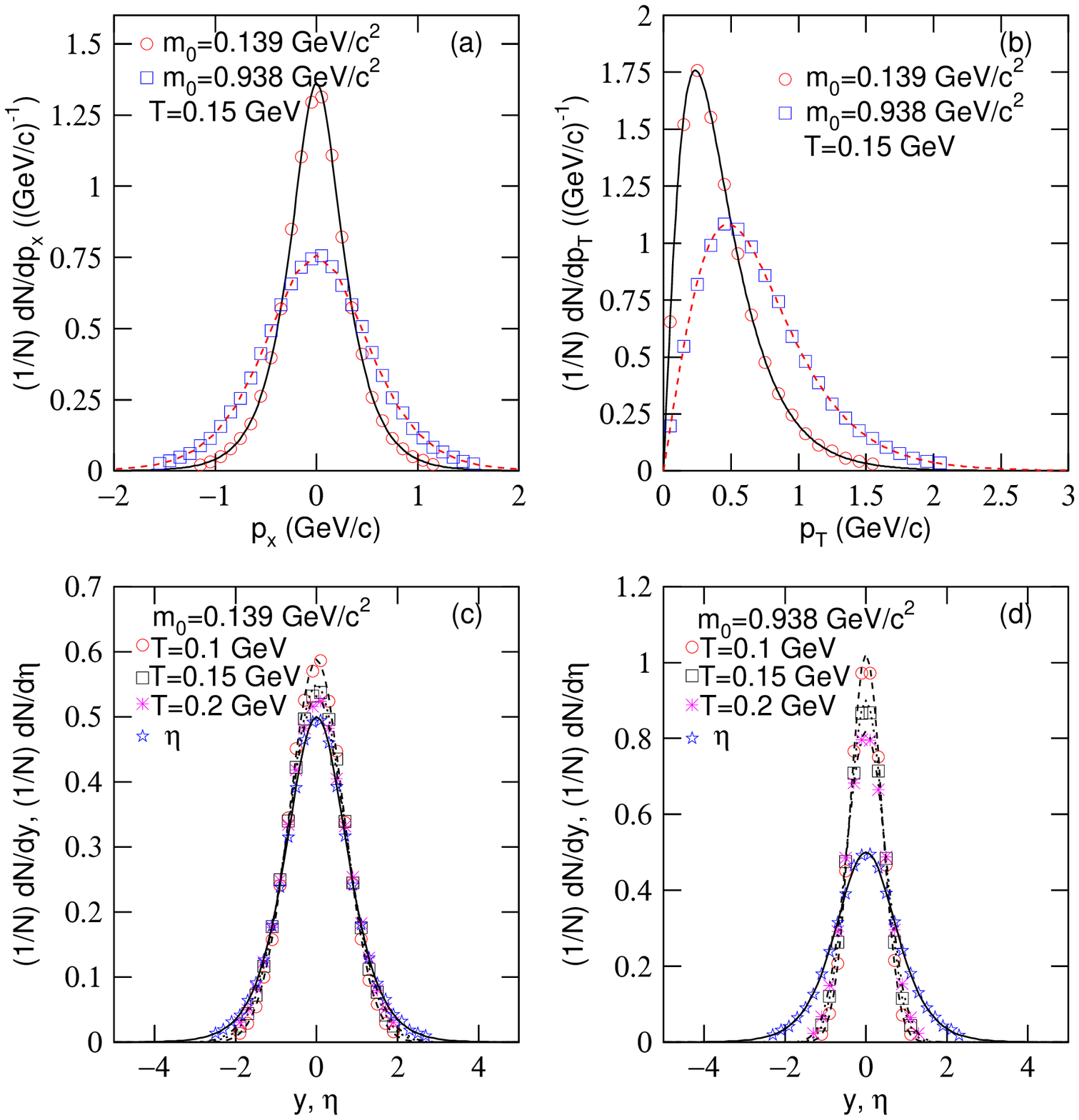}
\end{center}
{\small Fig. 4. Distributions of (a) $p_x$ for pions and protons
at $T=0.15$ GeV, (b) $p_T$ for pions and protons at given $T$, (c)
$y$ ($\eta$) for pions at three $T$ shown in the panel, and (d)
$y$ ($\eta$) for protons at three $T$, respectively, in the
framework of Tsallis statistics with $q=1.05$. The curves and
symbols represent the results obtained from analytical and Monte
Carlo calculations.}
\end{figure*}

The transverse momentum spectra obtained from the Hagedorn
function and its modified forms are presented in Fig. 1. The
solid, dashed, dotted, and dot-dashed curves represent the results
from Eqs. (1)--(4), respectively, with $p_1=p_2=p_3=p_4=10$
GeV/$c$ and $n_1=n_2=n_3=n_4=5$. In particular, the dashed curves
with marks I and II corresponding to $m_0=0.139$ (pion) and 0.938
GeV/$c^2$ (proton) in Eq. (2) respectively. One can see that the
effect of rest mass in the first modified form is very small due
to large $p_T$. The other two modified forms describe narrow $p_T$
range with high probability density in low $p_T$ region. In
particular, the third modified form has the maximum probability
density at $p_T=0$, which is not correct comparing with general
experimental data.

The transverse momentum spectra obtained from the simplest
standard distribution and its modified forms are shown in Fig. 2.
The solid, dashed, and dotted curves represent the results from
Eqs. (5)--(7), respectively, with $T_1=T_2=T_3=0.2$ GeV, $\mu=0.1$
GeV, $y_{\min}=-0.5$, and $y_{\max}=0.5$. The upper [(a)(b)],
middle [(c)(d)], and lower [(e)(f)] panels correspond to $S=1$, 0,
and $-1$, respectively; and the left [(a)(c)(e)] and right
[(b)(d)(f)] panels correspond to $m_0=0.139$ and 0.938 GeV/$c^2$
respectively. One can see that the modified forms contribute a
wider $p_T$ range than the simplest standard distribution, though
a lower probability density in very-low $p_T$ region in the
modified forms appears due to the limitation of normalization. For
pion $p_T$ distribution, the effect of $S$ is obvious, which
should be considered in the calculation due to small mass. For
proton $p_T$ distribution, the effect of $S$ is very small, which
can be neglected in the calculation due to large mass.

The transverse momentum spectra obtained from the simplest
Boltzmann distribution and its modified forms are displayed in
Fig. 3. The solid, dashed, and dotted curves represent the results
from Eqs. (8)--(10), respectively, with $T_1=T_2=T_3=0.2$ GeV. The
left [(a)] and and right [(b)] panels correspond to $m_0=0.139$
and 0.938 GeV/$c^2$ respectively. One can see that the modified
forms contribute a wider $p_T$ range than the simplest Boltzmann
distribution, though a lower probability density in very-low $p_T$
region in the modified forms appears due to the limitation of
normalization. For pion $p_T$ distribution, the effect of mass on
the two modified forms is not obvious, which does not need to be
distinguished clearly in the calculation. For proton $p_T$
distribution, the effect of mass on the two modified forms is
obvious, which should be distinguished in the calculation.

In the framework of Tsallis statistics with $q=1.05$, Figure 4
shows the distributions of (a) $p_x$ for pions and protons at
$T=0.15$ GeV, (b) $p_T$ for pions and protons at given $T$, (c)
$y$ ($\eta$) for pions at three $T$ shown in the panel, and (d)
$y$ ($\eta$) for protons at three $T$, respectively. The curves
represent the results calculated from Eqs. (25), (23), (24), and
(19), respectively, in the analytical calculation. The symbols
represent the results obtained from Eqs. (12), (27), and (17) in
the Monte Carlo calculation. One can see the natural result that
the analytical and Monte Carlo calculations are consistent with
each other. This also confirms that our calculations are correct.
Another observation is that the distribution of $y$ is closer to
that of $\eta$ at higher $T$, in particular for lighter particle.

From the above discussions one can see that the trends of modified
functions show large departures from that of original function in
some cases. Because of the limitation of normalization, the
increase (decrease) of probability in low $p_T$ region results in
the decrease (increase) of probability in very-low $p_T$ region.
The modified functions, Eqs. (2) and (3), do not cause large
departure from the original function, Eq. (1). The modified
function, Eq. (4), shows largely complete difference from the
original function.

The modified functions, Eqs. (6) and (7) [Eqs. (9) and (10)],
result in larger probability in low $p_T$ region and smaller
probability in very-low $p_T$ region comparing with the original
function, Eq. (5) [Eq. (8)]. In particular, for a given $p_T$
spectrum, the modified functions, Eqs. (6) and (7) [Eqs. (9) and
(10)], ``measure" lower temperatures than the original function,
Eq. (5) [Eq. (8)]. For example, if we use the modified functions
to ``measure" (fit) the spectra (solid curves) of original
function in Figs. 2 and 3, the modified temperatures for the
spectra of pions and protons are smaller than 0.2 GeV which is the
temperature of the spectra of original function. Contrarily, if we
use the original function to ``measure" the spectra (dashed and
dotted curves) of modified functions in Figs. 2 and 3, the
original temperatures for the spectra of pions and protons are
greater than 0.2 GeV which is the temperature of the spectra of
modified functions.

Figures 5 and 6 show the situations of the modified functions
``measuring" the spectra (solid curves) of the original one in
Figs. 2 and 3 respectively. The values of related temperature
parameters are shown in each panel. Other parameters for Fig. 5
are the same as for Fig. 2. One can see that the modified
temperatures ($T_2$ and $T_3$) for the spectra of pions and
protons are smaller than the original temperature ($T_1$), though
the modified functions does not fit the original one. This
inconsistent results render that the modified functions may be
necessary. Contrarily, in the case of the modified functions
fitting the original one, one can obtain consistent results which
mean that the modified functions are not necessary.

\begin{figure*}
\begin{center}
\includegraphics[width=12cm]{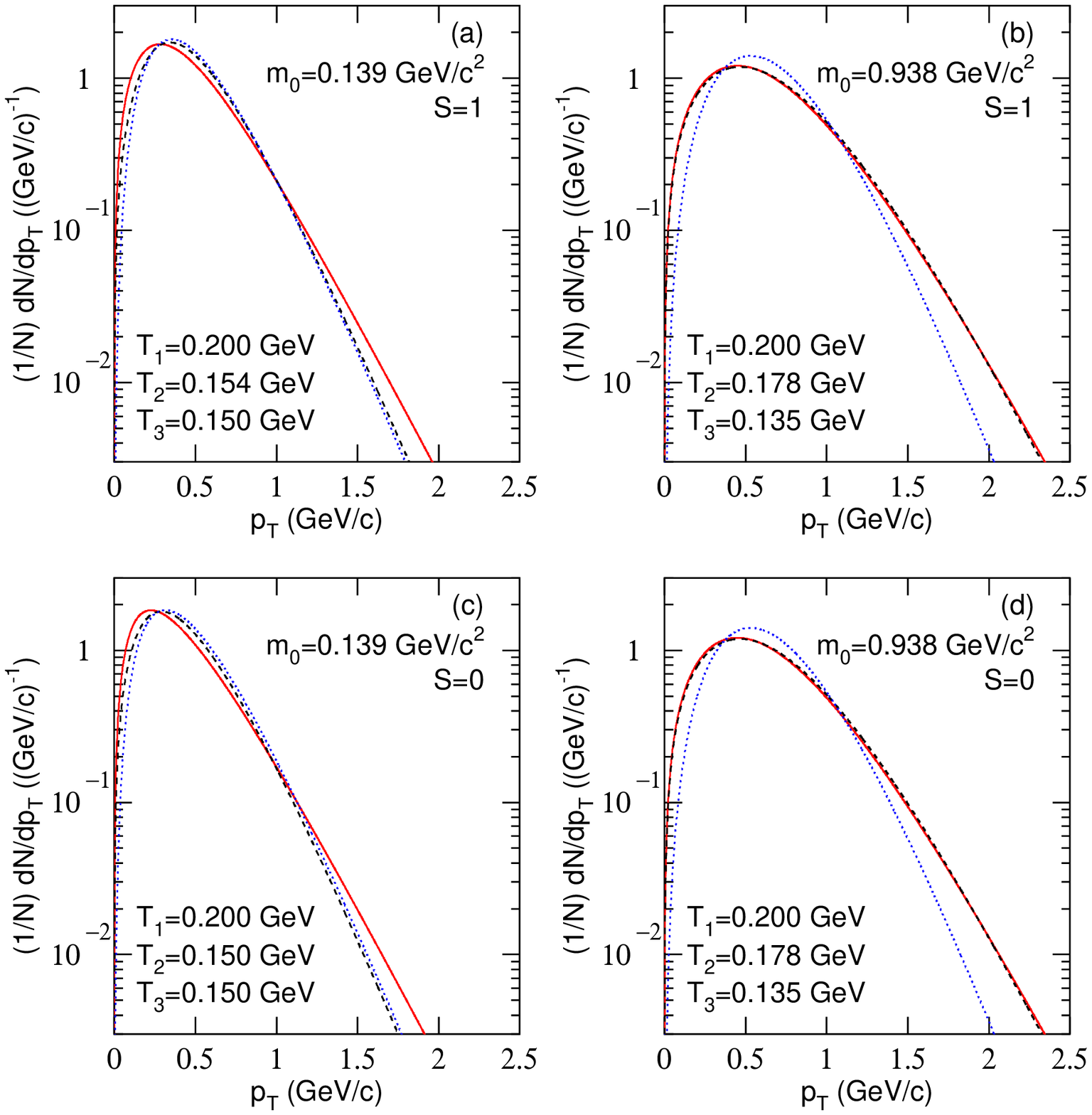}
\vskip.35cm
\includegraphics[width=12cm]{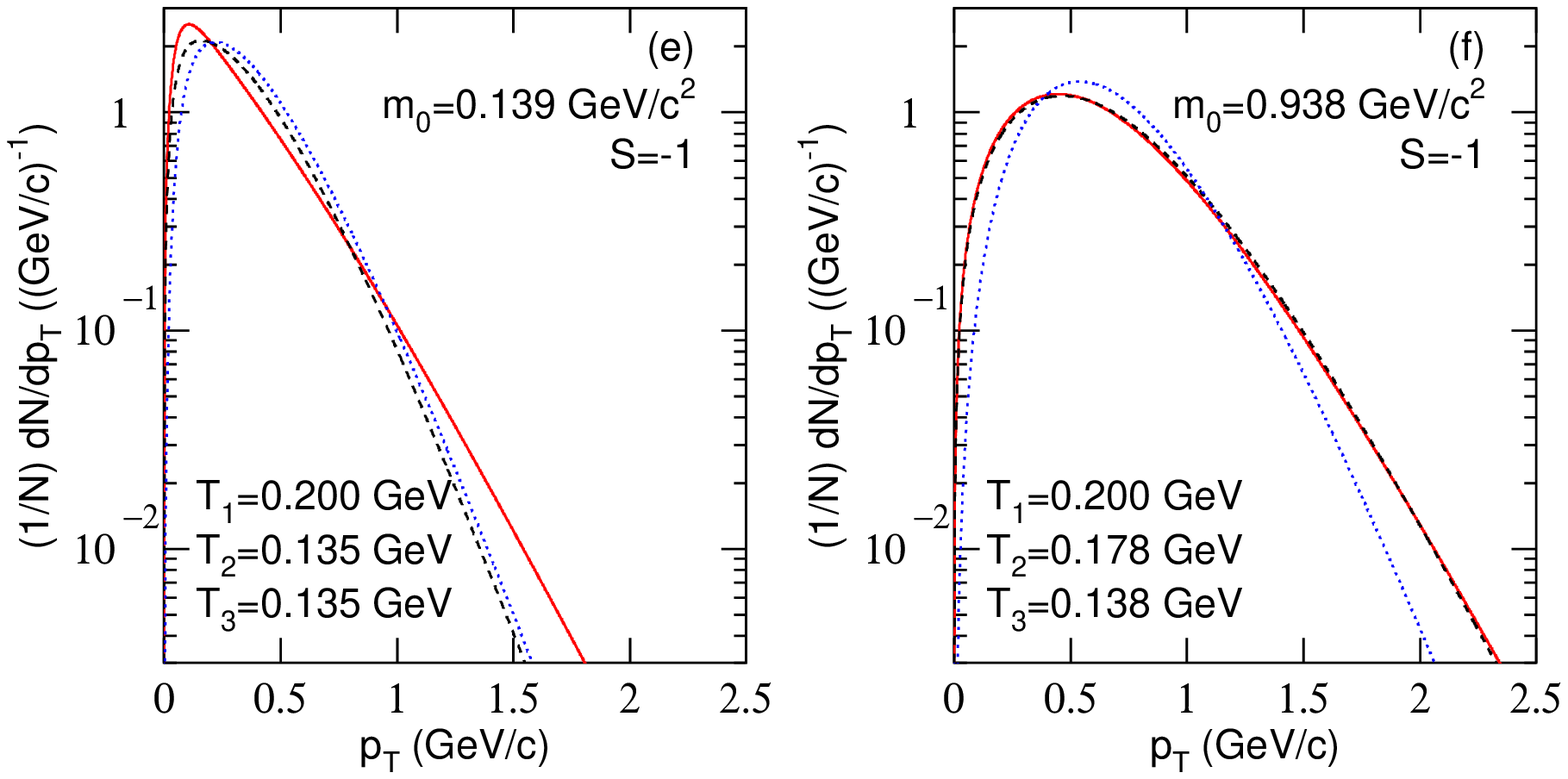}
\end{center}
{\small Fig. 5. Transverse momentum spectra obtained from the
simplest standard distribution and its modified forms which fit
the former. The solid, dashed, and dotted curves represent the
results from Eqs. (5)--(7) with $T_1$, $T_2$, and $T_3$,
respectively. Meanwhile, $\mu=0.1$ GeV, $y_{\min}=-0.5$, and
$y_{\max}=0.5$. Panels (a)(b), (c)(d), and (e)(f) correspond to
$S=1$, 0, and $-1$, respectively; and panels (a)(c)(e) and
(b)(d)(f) correspond to $m_0=0.139$ and 0.938 GeV/$c^2$
respectively.}
\end{figure*}

\begin{figure*}
\begin{center}
\includegraphics[width=12cm]{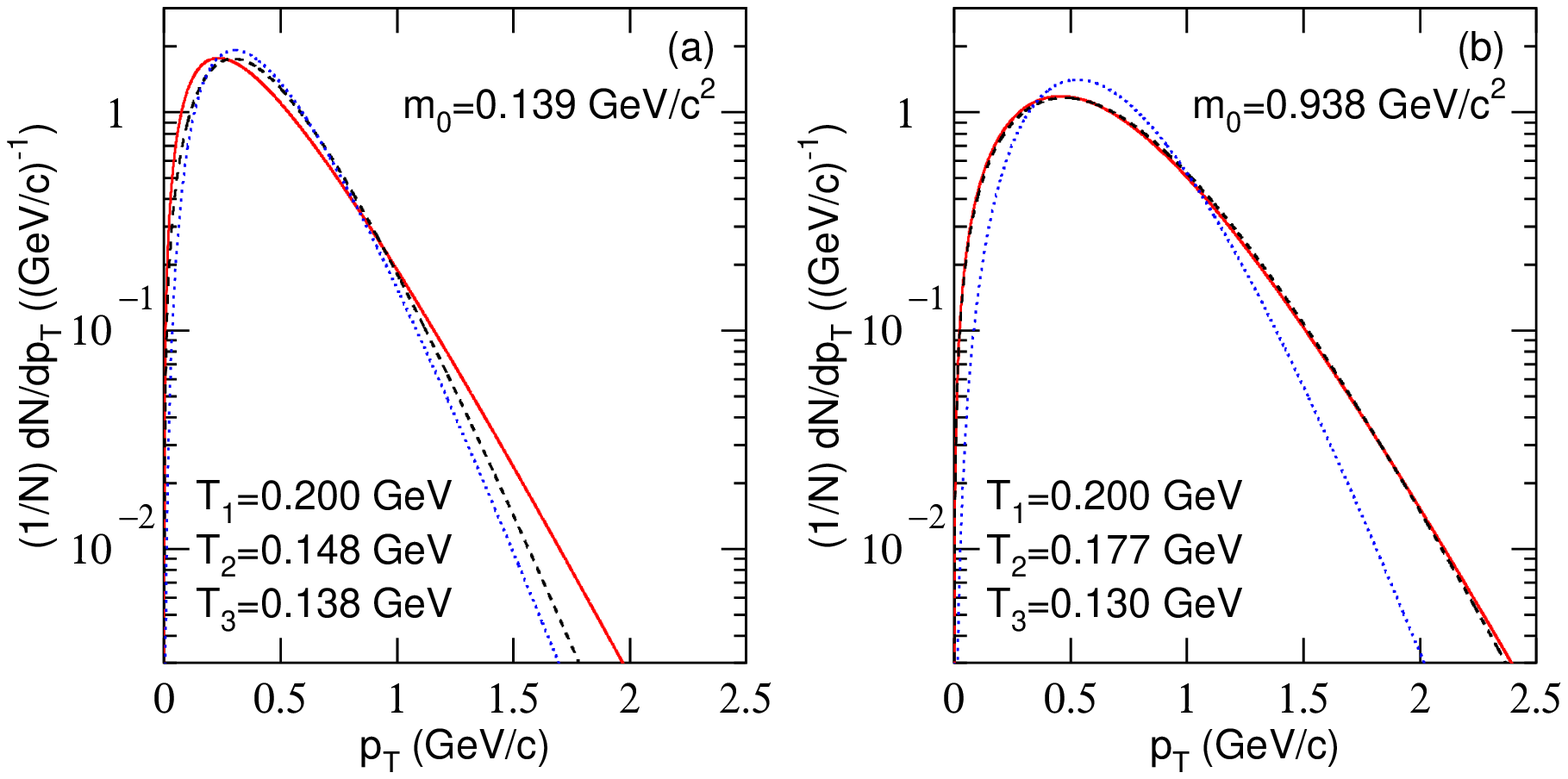}
\end{center}
{\small Fig. 6. Transverse momentum spectra obtained from the
simplest Boltzmann distribution and its modified forms which fit
the former. The solid, dashed, and dotted curves represent the
results from Eqs. (8)--(10) with $T_1$, $T_2$, and $T_3$,
respectively. Panels (a) and (b) correspond to $m_0=0.139$ and
0.938 GeV/$c^2$ respectively.}
\end{figure*}

\begin{figure*}
\begin{center}
\includegraphics[width=12cm]{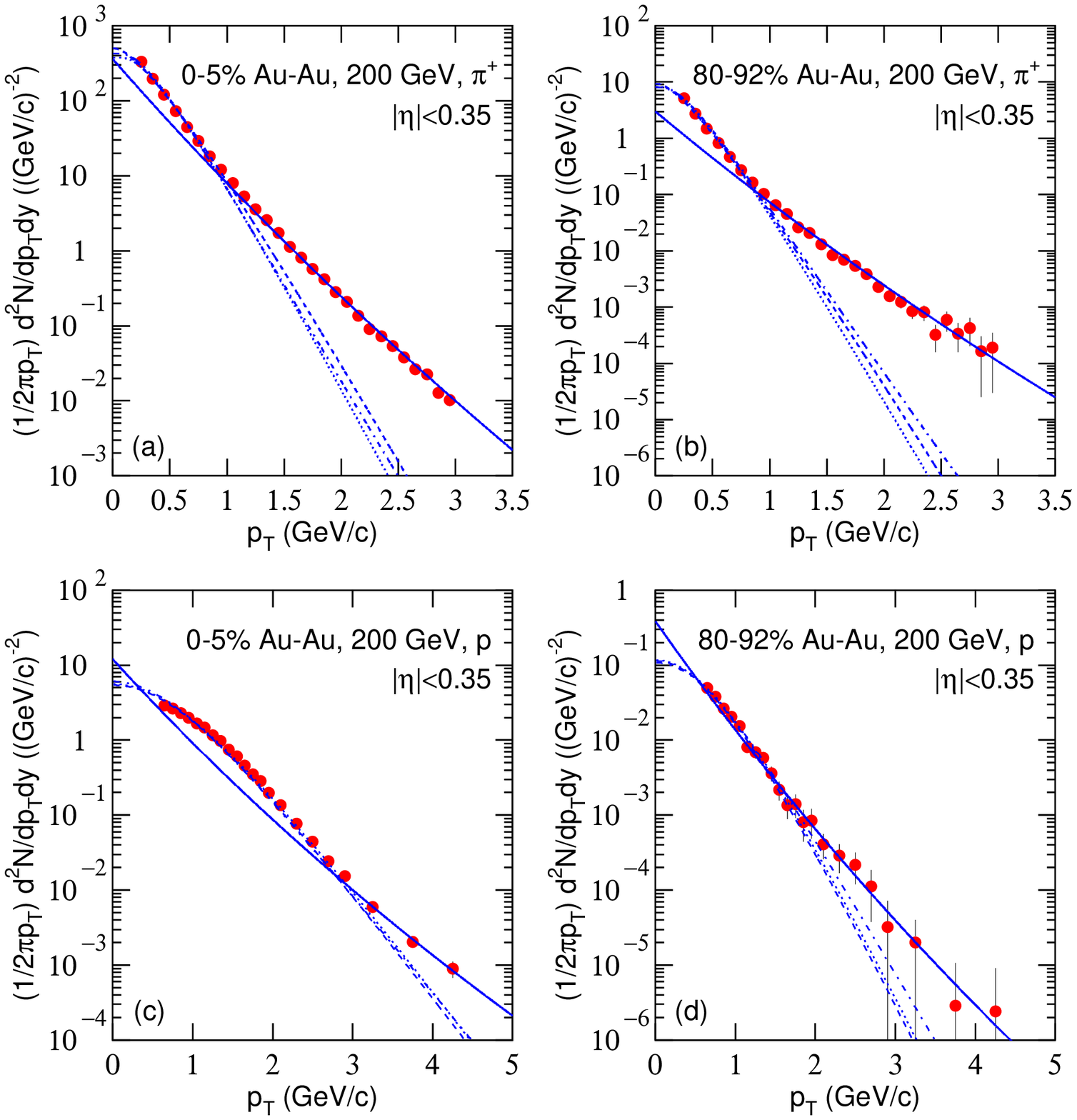}
\end{center}
{\small Fig. 7. Transverse momentum spectra of (a)(b) $\pi^+$ and
(c)(d) $p$ produced in (a)(c) 0--5\% and (b)(d) 80--92\% Au-Au
collisions at 200 GeV. The symbols represent the experimental data
measured by the PHENIX Collaboration [19]. The solid, dashed,
dotted, and dot-dashed curves represent the fitting results by
Eqs. (1), (6), (9), and (23), respectively.}
\end{figure*}

At the same temperature $T$, the distributions of $p_x$ and $p_T$
for pions are much narrower than those for protons due to the fact
that the distribution width increases with the increase of mass as
indicated in the ideal gas model based on the Maxwell-Boltzmann
statistics. With the increase of $T$, the distribution of $y$ is
closer to that of $\eta$ for not only pions but also protons. The
degree of closeness for pion spectrum is much more than that for
proton spectrum at a given temperature. This is a natural
conclusion due to the definitions of $y$ and $\eta$, though we
obtain this conclusion in the framework of Tsallis statistics and
the assumption of isotropic emission by the analytical and Monte
Carlo calculations.

Before conclusions, as an example of the applications of the above
distributions, Figure 7 present some comparisons with the $p_T$
spectra of (a)(b) positive pions ($\pi^+$) and (c)(d) protons
($p$) produced in (a)(c) central (0--5\%) and (b)(d) peripheral
(80--92\%) gold-gold (Au-Au) collisions at 200 GeV. The symbols
represent the experimental data measured by the PHENIX
Collaboration [19]. The solid, dashed, dotted, and dot-dashed
curves represent the fitting results by Eqs. (1), (6), (9), and
(23), respectively. In the calculation, from panels (a) to (d), we
take in proper order $n_1=40$, 39, 27, and 35 in Eq. (1) with
$p_1=10$ GeV/$c$; $T_2=0.16$, 0.13, 0.29, and 0.19 GeV in Eq. (6)
with $\mu=0$ at high energy and $y_{\min}\approx-0.35$ and
$y_{\max}\approx0.35$ in the experiment; $T_2=0.15$, 0.12, 0.29,
and 0.19 GeV in Eq. (9); $T=0.14$, 0.12, 0.27, and 0.19 GeV in Eq.
(23) with $q=1.01$. One can see that the mentioned distributions
describe partly the spectra of $\pi^+$ and $p$ produced in central
and peripheral Au-Au collisions at 200 GeV. The temperature in
central collisions is larger than that in peripheral collisions.
The temperature for the spectra of pions is smaller than that for
the spectra of protons.

Generally, Eq. (1) describes the spectra in high $p_T$ region due
to the hard scattering process. Eqs. (6), (9), and (23) describes
the spectra in low $p_T$ region due to the soft excitation
process. In particular, Eqs. (6), (9), and (23) are harmonious in
thermodynamics. To describe the spectra in whole $p_T$ region, a
superposition of Eq. (1) and one of Eqs. (6), (9), and (23) should
be used. There are two types of superpositions,
\begin{align}
f_0(p_T)=\frac{1}{N} \frac{dN}{dp_T}=kf_S(p_T)+(1-k)f_H(p_T)
\end{align}
and
\begin{align}
f_0(p_T)=& \frac{1}{N} \frac{dN}{dp_T} \nonumber\\
=& A_1\theta(p_1-p_T) f_S(p_T) + A_2 \theta(p_T-p_1)f_H(p_T),
\end{align}
where $f_S(p_T)$ denotes one of the soft components, Eqs. (6),
(9), and (23); $f_H(p_T)$ denotes the hard component, Eq. (1); $k$
denotes the contribution fraction of the soft component in Eq.
(28); $A_1$ and $A_2$ are constants which result in the two
components to be equal to each other at $p_T=p_1$; and
$\theta(p_1-p_T)$ and $\theta(p_T-p_1)$ are the usual step
function.

In our recent works [20, 21], to extract the kinetic freeze-out
temperature and transverse flow velocity, the two types of
superpositions are used respectively, where the soft component is
described by the blast-wave model with Boltzmann-Gibbs statistics
[22--24] and with Tsallis statistics [25--27]. There are small
differences ($<5\%$) in the parameters extracted by the two
superpositions. The first superposition can obtain a smooth curve
easily, and the parameters are entangled in the extraction
process. The second superposition has no entanglement in the
extraction process of the parameters, and the curves are possibly
not smooth in the point of split joint, $p_1$.
\\

{\section{Conclusions}}

To conclude, the transverse momentum spectra obtained from
different functions or distributions are compared. For the
Hagedorn function, the effect of rest mass in the first modified
form is very small due to large transverse momentum. The other two
modified forms describe narrow transverse momentum range. For the
simplest standard and Boltzmann distributions, the modified forms
contribute a wider transverse momentum range than the original
distributions, though a lower probability density in very-low
transverse momentum region appears in the modified forms.

For a given transverse momentum spectrum, the modified forms
``measure" lower temperature comparing with the simplest standard
and Boltzmann distributions. Comparing with the original function
with its modified forms, it is hard to say that which one is
better. Based on the Tsallis momentum distribution and the
isotropic assumption, the distributions of momentum components,
transverse momenta, rapidities, and pseudorapidities for pions and
protons are obtained by the analytical and Monte Carlo methods. It
is natural that the rapidity distribution is closer to the
pseudorapidity one at higher temperature and with smaller mass.

Comparing with the experimental data measured by the PHENIX
Collaboration, Eq. (1) is confirmed to fit the spectra in high
transverse momentum region. As the harmonious distributions in
thermodynamics, Eqs. (6), (9), and (23) are confirmed to fit the
spectra in low transverse momentum region. To fit the spectra in
whole transverse momentum region, two types of superpositions,
Eqs. (28) and (29), which combine Eq. (1) and one of Eqs. (6),
(9), and (23) are suitable. In the superpositions, Eqs. (2)--(4)
can replace Eq. (1), Eqs. (5) and (7) can replace Eq. (6), and
Eqs. (8) and (10) can replace Eq. (9).
\\

{\bf Data Availability}

The data used to support the findings of this study are quoted
from the mentioned references. As a phenomenological work, this
paper does not report new data.
\\

{\bf Conflicts of Interest}

The authors declare that there are no conflicts of interest
regarding the publication of this paper.
\\

{\bf Acknowledgments}

This work was supported by the National Natural Science Foundation
of China under Grant Nos. 11575103 and 11747319, the Shanxi
Provincial Natural Science Foundation under Grant No.
201701D121005, and the Fund for Shanxi ``1331 Project" Key
Subjects Construction.
\\

\end{multicols}
\end{document}